\documentclass[prl,twocolumn,showpacs,amsmath,amssymb,superscriptaddress]{revtex4}

\usepackage[dvips]{graphicx}
\usepackage{graphics}
\usepackage{dcolumn}
\usepackage{bm}

\begin{document}

\title{\begin{center}Tunneling Anisotropic Magnetoresistance: A spin-valve like tunnel magnetoresistance using a single magnetic layer\end{center}}

\author{C. Gould}
\affiliation{Physikalisches Institut (EP3), Universit\"{a}t
W\"{u}rzburg, Am Hubland, D-97074 W\"{u}rzburg, Germany}

\author{C. R\"{u}ster}
\affiliation{Physikalisches Institut (EP3), Universit\"{a}t
W\"{u}rzburg, Am Hubland, D-97074 W\"{u}rzburg, Germany}

\author{T. Jungwirth}
\affiliation{Institute of Physics ASCR, Cukrovarnická 10, 162 53
Praha 6, Czech Republic} \affiliation{School of Physics and
Astronomy, University of Nottingham, Nottingham NG7 2RD, UK}
\affiliation{Department of Physics, University  of Texas, Austin
TX 78712-0264}

\author{E. Girgis}
\affiliation{Physikalisches Institut (EP3), Universit\"{a}t
W\"{u}rzburg, Am Hubland, D-97074 W\"{u}rzburg, Germany}

\author{G. M. Schott}
\affiliation{Physikalisches Institut (EP3), Universit\"{a}t
W\"{u}rzburg, Am Hubland, D-97074 W\"{u}rzburg, Germany}

\author{R. Giraud}
\affiliation{Physikalisches Institut (EP3), Universit\"{a}t
W\"{u}rzburg, Am Hubland, D-97074 W\"{u}rzburg, Germany}

\author{K. Brunner}
\affiliation{Physikalisches Institut (EP3),
Universit\"{a}t W\"{u}rzburg, Am Hubland, D-97074 W\"{u}rzburg,
Germany}

\author{G. Schmidt}
\affiliation{Physikalisches Institut (EP3), Universit\"{a}t
W\"{u}rzburg, Am Hubland, D-97074 W\"{u}rzburg, Germany}

\author{L.W. Molenkamp}
\affiliation{Physikalisches Institut (EP3), Universit\"{a}t
W\"{u}rzburg, Am Hubland, D-97074 W\"{u}rzburg, Germany}

\date{\today}

\begin{abstract}
We introduce a new class of spintronics devices in which a spin-valve like effect results from strong spin-orbit coupling in a
single ferromagnetic layer rather than from injection and
detection of a spin-polarized current by two coupled ferromagnets.
The effect is observed in a
normal-metal/insulator/ferromagnetic-semiconductor tunneling
device. This behavior is caused by the interplay of the
anisotropic density of states in (Ga,Mn)As with respect to the
magnetization direction, and the two-step magnetization reversal
process in this material.

\end{abstract}


\maketitle

Devices relying on spin manipulation are hoped to provide
low-dissipative alternatives for microelectronics. Furthermore, spintronics
is expected to lead to
full integration of information processing and storage
functionalities opening attractive prospects for the realization
of instant on-and-off computers. A primary goal of
current spintronics research is to realize a device with metal
spin-valve like behavior \cite{Moodera} in an all
semiconductor-based structure enhancing
integration of spintronics with existing microelectronics
technologies. An oft proposed scheme for such a device
consists of a tunnel barrier between two ferromagnetic
semiconductors. As such, (Ga,Mn)As/(Al,Ga)As/(Ga,Mn)As structures
have previously been studied \cite{Higo,ohnonature04} with some
promising results. However, realizing the full potential of these
systems will require a complete understanding of the physics of
tunneling into (Ga,Mn)As, which we have found to be rather
different than previously thought.

In this spirit, we investigate transport in a structure
consisting of a single ferromagnetic (Ga,Mn)As layer fitted with a
tunnel barrier and a non-magnetic metal contact. We report some of the rich experimental properties
of such a tunneling structure and provide an
interpretation of the measured spin-valve like effect as a
tunneling anisotropic magnetoresistance (TAMR) due to a two-step
magnetization reversal and a magnetization dependent density of states 
(DOS) in the (Ga,Mn)As layer.

The magnetic layer in our sample is a 70 nm thick epitaxial
(Ga,Mn)As film grown by low temperature (270~$^\circ$C) molecular
beam epitaxy onto a GaAs (001) substrate \cite{Shen}.
High-resolution x-ray diffraction showed that the sample had high
crystalline quality comparable to that of the substrate. From the
measured lattice constant and the calibration curves of
Ref.~\cite{Schott}, the Mn concentration in the ferromagnetic
layer is roughly 6\%. Etch capacitance-voltage control
measurements yielded a hole density estimate of
$\sim10^{21}$~cm$^{-3}$ and the Curie temperature of 70~K was
determined from SQUID measurements.

After growth, the sample surface was Ar sputtered to remove any potential oxides,
and a 1.4~nm Al layer was deposited at a rate of 0.4 \AA /sec and a base pressure of $2\times
10^{-6}$mbar using Ar gas. The Al layer was oxidized in-situ using
100 mbar of pure oxygen for 8 hours, producing a closed AlOx layer
and thereby forming a tunnel barrier. An electrical contact was
then fashioned onto the structure by evaporating 5~nm of Ti as a
sticking layer followed by 300~nm of Au.
\begin{figure}[h]
\hspace*{-0cm}
\includegraphics[angle=0,width=8cm]{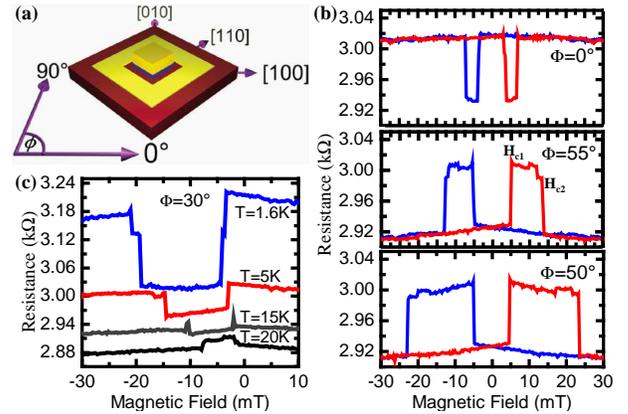}

\vspace*{-0cm}
  \caption{a) Device schematic showing the contact geometry
and the crystallographic directions. b) Hysteretic magnetoresistance curves acquired at 4.2K with 1 mV bias by sweeping the magnetic
  field along the 0$^\circ$, 50$^\circ$, and 55$^\circ$ directions. Spin valve like features of varying widths and signs
  are clearly visible, delimited by two switching events labeled $H_{c1}$ and $H_{c2}$. The magnetoresistance  is independent of the bias direction or amplitudes up to 1 meV.
  c) TAMR along 30$^\circ$ for temperatures from 1.6 K to 20 K, showing a change of sign of the signal. The curves are vertically offset for clarity.}
  \label{figure1}
\end{figure}
Standard optical lithography and chemically assisted ion beam
etching (CAIBE) were then used to pattern the device as shown in
Fig.~\ref{figure1}a. In the first step, material is etched away,
leaving only the central 100 x 100 $\mu$m$^{2}$ square pillar
consisting of the metal contact on a tunnel barrier. The
surrounding W sticking layer and Au contact are then deposited onto
the (Ga,Mn)As surface, providing a back contact.

The (Ga,Mn)As resistivity is $1.1\times 10^{-3}
~\Omega$cm, typical for high quality material
\cite{Edmonds}, and corresponding to a resistance of $\sim 10 ~\Omega$ between the central 
pillar and the backside contact. This was confirmed by measuring the resistance through
similar pillars without a tunnel barrier. This resistance is over
two orders of magnitude lower than that of the total device,
rendering any bulk magnetoresistance of the (Ga,Mn)As negligible.

The sample was inserted into a magnetocryostat allowing for the application of magnetic fields of
up to 300 mT in any direction. Results discussed here are for
fields in the plane of the magnetic layer with the field direction given by
its angle $\phi$ with respect to the [100]
direction, as indicated in Fig.~\ref{figure1}a.

Fig.~\ref{figure1}b presents representative magnetoresistance curves at various
angles. For each curve, the field is swept from negative
saturation to positive saturation and back, but the plot
focuses on the interesting region from -30 to +30 mT. In all
cases, the magnetoresistance shows spin-valve like behavior
with an amplitude of $\sim 3\%$ delimited by two switching events
(labelled $H_{c1}$ and $H_{c2}$ in the figure) between which the
resistance of the sample is different from its value outside these
events. However, the width and even the sign of the TAMR feature
depend on $\phi$. In comparing the curves of
Fig.~\ref{figure1}b, we emphasize that despite the feature
changing signs as a function of $\phi$, the
device appears to have only two distinct resistance states; a low
one of $\sim 2920~\Omega$ and a high one of $\sim 3000~\Omega$.

\begin{figure}[b!]

\includegraphics[angle=0,width=8.5cm]{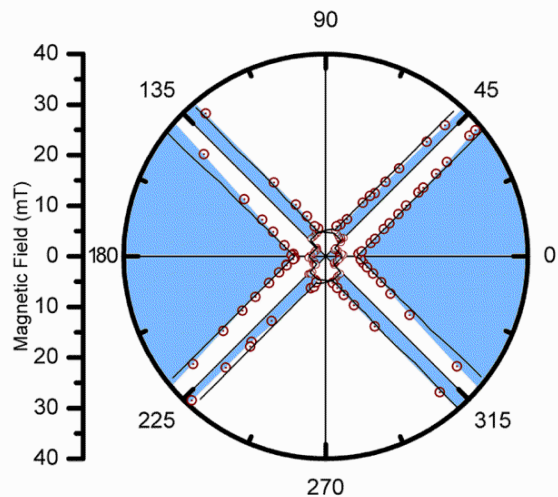}

  \caption{Polar plot compiled from individual magnetoresistance
  curves. The circles indicate
  the switching events $H_{c1}$ and $H_{c2}$ from the individual curves.
  The shaded areas are regions where the sample is in a high
  resistance state. The solid lines are a fit to the model described in the text.}
  \label{figure2}
\end{figure}

In order to better understand this behavior, we
summarize the data from field sweeps at many angles in the polar
plot of Fig.~\ref{figure2}. Here the open circles represent the
fields at which the switching events $H_{c1}$ and $H_{c2}$ occur
in the individual sweeps. These delimit boundaries
between sections of higher and lower resistance. Shaded areas
indicate regions where the sample is in its high resistance
state. Viewed in this way, the loci of switching events form a
highly symmetric pattern with a striking resemblance to
switching previously observed in magneto-optical studies of
epitaxial Fe films \cite{Cowburn} and (Ga,Mn)As \cite{Moore} as
well as in transport studies on (Ga,Mn)As in the in-plane Hall
geometry \cite{Tang}, and associated with materials that reverse
their magnetization $M$ in two steps by the nucleation and
propagation of 90$^\circ$ domain walls.
Within single-domain theory, the expression for
the total magnetic energy $E_{m}$ of our system is:
\begin{equation}\label{equation1}
E_m=K_u sin^2(\theta)+K_c sin^2(2\theta)-MH cos(\theta-\phi),
\end{equation}
where $K_c$ is the cubic anisotropy known to be dominant in
(Ga,Mn)As \cite{Moore,Tang,Hrabovsky}, while $K_u$ is the
uniaxial anisotropy which is also often observed in (Ga,Mn)As
\cite{Moore}. $H$ is the amplitude of the applied magnetic field and
$\theta$ is the angle of the magnetization measured from the
[100] crystal direction.

Since the magnetization reversal takes place through domain
walls propagating through the structure, the picture of
Stoner-Wohlfarth \cite{Stoner} of a coherent magnetization
reversal does not apply (neglecting rotations away from the
cubic easy axis at higher $H$). Instead, as discussed in
Ref.~\cite{Cowburn}, the magnetization will switch from its local
minimum to the global energy minimum as long as the energy gained
in doing so is larger than the energy required to
nucleate/propagate a domain wall through the sample. Calling this
energy $\epsilon$, it follows from the form of $E_m$ that as $H$
is swept the switching of the magnetization can
take place in two steps. In the first step, $M$ switches from the
cubic easy axis closest to the initial direction of $H$ to a global
easy axis 90$^\circ$ askew from this
one. Then, in the second step, $M$ switches by an additional 90
degrees completing its reversal. Pursuing the analysis, one finds that the fields at which these switching
events take place are given by
$H_{c1,2}=(\epsilon \pm K_u)/(M | |\cos(\phi)| \pm |\sin(\phi)|
|)$,
where the plus (minus) sign in the denominator is for $H_{c1}$ ($H_{c2}$).
The sign before $K_u$ depends on
if the switching is towards or away from a uniaxial easy
axis. The sign therefore reverses every 90 degrees and is opposite
for $H_{c1}$ and $H_{c2}$ \cite{Cowburn}. Fitting the above equation to our data produces the solid line in
the polar plot of Fig.~\ref{figure2}. This yields a value of $450$ 
erg/cm$^{3}$ for $K_{u}$ and $1550$ erg/cm$^{3}$ for $\epsilon$. We
confirmed the two step switching behavior of the sample through
SQUID measurements.

From this analysis and Fig.~\ref{figure2} it is clear that our
sample is in a high resistance state when $M$ lies along the [100]
or [\=100] crystallographic direction, and has a lower resistance
when $M$ is along [010] or [0\=10].
 This picture is further supported by the behavior of
the magnetoresistance at higher $H$. When the magnetic
field is not aligned along an easy axis, and the field is swept to
full saturation, the magnetization will rotate away from the easy
axis to the direction parallel to the field. A corresponding gradual
change in resistance is then observed consistent with a cubic anisotropy
at least an order of magnitude larger than $K_u$, in agreement with Ref.~\cite{Tang}.

We now turn to a theoretical analysis illustrating that
anisotropies in the (Ga,Mn)As DOS with respect to the
magnetization orientation are large enough to explain the
observation of the TAMR effect. The electronic structure of the
(Ga,Mn)As is calculated using the $\vec{k}\cdot \vec{p}$ envelope
function description of the GaAs host valence bands in the
presence of an effective exchange field, $\vec{h}=J_{pd}
\vec{S}_{Mn}$, produced by the polarized Mn local moments with
spin density $\vec{S}_{Mn}$ \cite{Abolfath}. The broken
in-plane cubic symmetry responsible for the difference between
tunnel resistances for $M$ along [100] and [010] is
theoretically modelled by introducing an in-plane uniaxial strain
of order 0.1\%. Due to a very strong spin-orbit interaction in the
valence band, such a small strain leads to values of $K_u$
comparable to the one estimated above and also to sizable DOS
anisotropies.

\begin{figure}[h]

\vspace*{0cm}
  \includegraphics[width=3in]{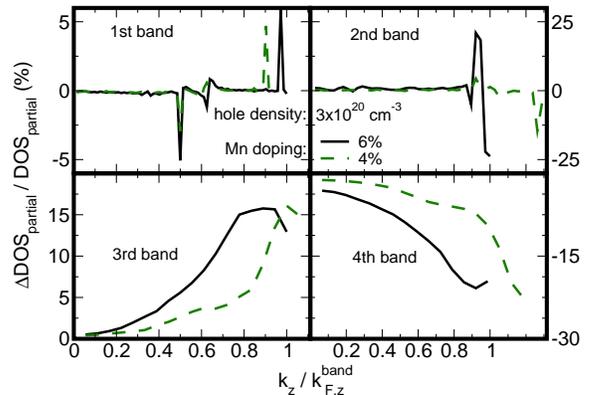}

\vspace*{-.5cm}
  \caption{The relative difference between partial DOS at the Fermi energy for
$M$ along [010] and [100] directions is plotted separately for
each of the four occupied valence bands.  Note that in the ferromagnetic state, even the near k=0, states cannot properly be called light and heavy holes due to the p-d exchange interaction. Dashed lines
correspond to Mn$_{\rm Ga}$ concentration of 4\%; solid lines
corresponding to 6\% Mn doping are shown for comparison.}
  \label{figure3}

\end{figure}

Defining the partial DOS as the DOS at a given k$_z$ and for a
given band, we show in Fig.~\ref{figure3} the relative partial DOS
anisotropy ($\Delta {\rm DOS_{partial}} \equiv {\rm DOS_{partial}}
(M\|[010]) - {\rm DOS_{partial}} (M\|[100])$) at the Fermi energy $E_F$
calculated as a function of the out-of-plane wavevector $k_z$ for
each of the four occupied bands that derive from the GaAs heavy-
and light-hole states which are spin-split due to the presence of the
Mn-moment induced exchange field ~\cite{Abolfath}. $k_{F,z}^{band}$ is the Fermi
wavevector in the given band for Mn$_{\rm Ga}$ concentration of
6\%.

Note that the experimental Curie temperature of 70 K is reproduced
theoretically assuming the hole density $3\times
10^{20}$~cm$^{-3}$ and 4\% of the cation sites occupied by Mn,
which is reasonably consistent with the experimental
estimates. The total DOS (${\rm DOS_{total}}$)
obtained by integrating over all $k_z$ up to the Fermi wavevector
$k_{F,z}$ and summing over all bands, has an anisotropy at $E_F$
of less than 1\% with respect to the magnetization
orientation. The tunnel conductance is, however, proportional to
the ${\rm DOS_{total}}$ only if in-plane momentum is not conserved
during the tunneling. For cleaner barriers and interfaces,
in-plane momentum is at least partially conserved resulting in,
roughly speaking, a higher probability of tunneling for states
with higher band and $k_z$ indices. As demonstrated in
Fig.~\ref{figure3}, the ${\rm DOS_{partial}}$ of these states can
change by tens of percent upon magnetization reorientation.
Fig.~\ref{figure3} also suggests that the magnitude and even the
sign of the overall tunnel magnetoresistance effect depend on
parameters of the (Ga,Mn)As film, such as the density of local
spins on substitutional Mn impurities, or on the barrier and
interface character which may select different ranges of band and
$k_z$ states that dominate the tunneling current.

\begin{figure}[h]

\vspace*{-0cm}
  \includegraphics[width=3in]{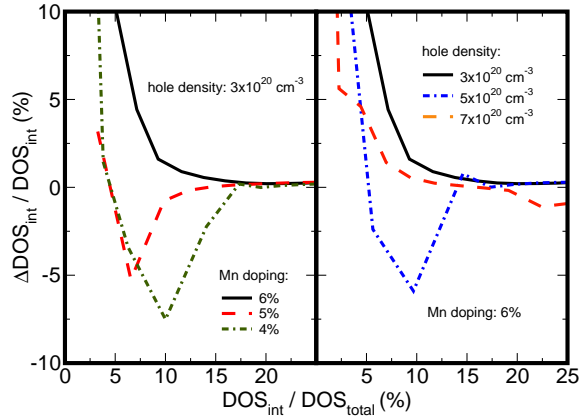}

\vspace*{-.1cm}
  \caption{The relative integrated DOS anisotropy
is plotted for different Mn (left panel) and hole (right panel)
concentrations. The x-axis represents the  DOS at the Fermi energy
that is assumed to contribute to tunneling, relative to the total
DOS at the Fermi energy. Moving from left to right corresponds to
gradually relaxing the momentum conservation condition. }
  \label{figure4}
\end{figure}

To estimate the overall size of the magnetoresistance effect
produced by the (Ga,Mn)As ${\rm DOS_{partial}}$ anisotropy we
start with the assumption that for clean barriers (perfect
in-plane momentum conservation) the tunneling is dominated by
states in the (Ga,Mn)As with $k_z$ close to $k_F$
in each band and that the tunneling probability of these states is
independent of the band index. We then gradually relax the
momentum conservation condition by adding states at $E_F$ with
decreasing $k_z$. In Fig.~\ref{figure4} we plot the
relative difference between this integrated ${\rm DOS_{int}}$
(integrated over the assumed range of $k_z$ contributing to
tunneling and summed over the four occupied bands) for the two
magnetization orientations. For $\sim$10\% of the total DOS at $E_F$
participating in the tunneling, the
theoretical ${\rm DOS_{int}}$ anisotropy is consistent
with the experimentally observed TAMR of order several percent.

The curves in the left panel of Fig.~\ref{figure4} are labelled by
different Mn doping concentrations and illustrate the general
dependence of the magnetoresistance effect on the Mn local spin
density. On a mean-field level this can be understood by recalling
that the (Ga,Mn)As electronic structure depends only on the
overall value of the effective exchange field $\vec{h}=J_{pd}
\vec{S}_{Mn}$, whether the spin-density magnitude $|S_{Mn}|$
changes through varying the number of Mn impurities at a fixed
temperature or through the temperature-dependent average spin
polarization of an individual Mn local moment at a fixed doping
level. The data in the left panel of Fig.~\ref{figure4} therefore
suggest that the sign of the TAMR can
change with temperature. We emphasize that this change in sign
occurs without a change in sign of the uniaxial anisotropy energy
constant. The right panel in Fig.~\ref{figure4} also predicts a strong
dependence of the TAMR on the number of holes in the (Ga,Mn)As
valence band.

In our sample the Mn doping and hole density are obviously fixed.
The temperature dependence, however, can be tested and the
experiment confirms the change of sign seen in the above
theoretical curves. Fig.~\ref{figure1}c shows a set of
magnetoresistance curves along 30$^\circ$ for temperatures from 1.6 to 20 K. At 1.6 K, the TAMR signal is negative.
Its amplitude gradually decreases to zero by 15 K, changes sign
and grows again as temperature is raised to 20 K. In fact, as temperature is increased from 4 K to 20 K, the entire
polar plot reverses signs. Finally, the TAMR disappears by 30 K, when
the magnetic anisotropy energy is no longer resolvable. 
Since the sign of $K_u$ does not change with temperature, this is
an experimental confirmation that the transport and magnetic
anisotropies can vary independently in our system.

The TAMR studied here shows a rich phenomenology that opens new
directions in spintronics research. Avoiding the second
ferromagnetic layer may have fundamental consequences for the
operation at high temperatures as it eliminates the need for a
buried ferromagnetic layer which cannot be effectively treated by
post-growth annealing \cite{ohnoapl03}. The data also demonstrate
that the sign of the spin-valve like signal, i.e., whether a
high- or low-resistance state is realized at saturation, can
change with the angle at which magnetic field is applied, with
temperature, or structural parameters of the (Ga,Mn)As layer,
interfaces, and the tunnel barrier.

Last but not least, our experiments provide a new perspective on
tunnel magnetoresistance in structures with two ferromagnetic
contacts. We demonstrate the need for caution in analyzing
spin-valve experiments, especially in materials where strong
spin-orbit coupling is present. As we have seen here, the
existence of a spin-valve like signal does not automatically imply
the injection and detection of a spin-polarized current in the
tunneling structure. Instead, two distinct material
properties combined in a constructive way can lead to bistable
magnetoresistive devices with unprecedented properties. We also
note that the amplitude of the effect discussed here may be 
optimized by using barriers with greater momentum conservation such as, for example, epitaxial AlAs.

The authors thank J. Sinova and A. H. MacDonald for
useful discussions and V. Hock for sample preparation, and
acknowledge financial support from the EC (SPINOSA), the German BMBF (13N8284) and DFG (SFB410), the
DARPA SpinS program and the Grant Agency of the Czech Republic
under grant 202/02/0912.

\end{document}